\documentclass[11pt]{article}
\usepackage{amsmath, amssymb}
\usepackage{slashed}
\usepackage{verbatim}
\usepackage{latexsym}
\usepackage{euscript}
\usepackage{amsfonts}
\usepackage{verbatim}
\usepackage[]{array}
\usepackage[]{mathrsfs}
\usepackage{graphicx}
\usepackage{amsmath}
\usepackage{amssymb}
\usepackage{amsxtra}
\usepackage{color}
\def\be{\begin{eqnarray}}
\def\ee{\end{eqnarray}}
\def\0{\nonumber}

\def\sfd{{\sf d}}
\usepackage{slashed}
\usepackage{verbatim}
\usepackage{latexsym}\usepackage{color}
\usepackage{euscript}

\newcommand\EA{\EuScript{A}}

\normalfont\large
\begin{document}
\vskip 2cm
\begin{center}
{\LARGE {\bf  Addendum: Weyl cocycles, (1986 {\it Class.Quantum Grav. {\bf 3} 635 }})}
\vskip 1cm

{\large  L.~Bonora
\\
\textit{ International School for Advanced Studies (SISSA),\\Via
Bonomea 265, 34136 Trieste, Italy   }
 }
\end{center}
\vskip 1cm
{\bf Abstract}. Weyl 0- and 1-cocycles of canonical dimension 6 in six dimensions, which were computed earlier in ref.\cite{bonorabregolapasti1986}, are recalculated from scratch. The analysis yields five Weyl invariants (0-cocycles), instead of four, and the same four non-trivial 1-cocycles (possible trace anomalies), like in that reference (up to the correction of one typo).

\vskip 1cm

\section{Introduction}

This is an Addendum to  ref.\cite{bonorabregolapasti1986}. That paper was a cohomological analysis, a method introduced in ref.\cite{bonoracottareina1983}, to calculate Weyl cocycles of canonical dimensions equal to the dimension of the spacetime, in dimensions 2, 4, and 6. In particular in 6 dimensions, four 0- and four 1-cocycles were found. Subsequently other similar or related papers appeared, \cite{deserschwimmer1993, bastianelli2001,bastianelli2001B,farhoudi2006,astaneh2015, beccaria2016,ferreira2017,coriano2018,ferreirashapiro2019,aros2021},  with not always overlapping results. The preparation of the book \cite{bonora2023} has prompted me to redo the calculations from scratch for the six-dimensional case by solving, in particular, the relative linear systems with Mathematica. The results differ from ref.\cite{bonorabregolapasti1986} in two respects: first, the Weyl invariants (0-cocycles) are five, not four; second, the 1-cocycles (the possible anomalies) are the same four, except for a different notational convention and one coefficient that has been corrected. Below these results are reported together with a good amount of details, which were understood in \cite{bonorabregolapasti1986}.

Before going to the explicit derivations I briefly recall the approach of \cite{bonoracottareina1983}. 
The basic Weyl transformation is 
\be
\delta_\omega g_{\mu\nu} =2 \omega g_{\mu\nu}\label{deltagmunu}
\ee
where $\omega$ is an arbitrary infinitesimal smooth function, with the corresponding functional operator (denoted with the same symbol)
\be
\delta_\omega 
= \int dx \, \delta_\omega g_{\mu\nu}
\frac {\delta}{\delta g_{\mu\nu}(x)}\label{deltaomega}
\ee
For simplicity we will adopt the anticommuting formalism, whereby $\omega$ is promoted to an anticommuting field. It follows that $\delta_\omega$ is nilpotent
\be
\delta_\omega^2=0\label{nilpotence}
\ee
A 0-cocycle is defined to be a local integral of the metric, its inverse and their derivatives of canonical dimension 6, $C[g]$, satisfying
\be
\delta_\omega C[g]=0\label{invariant}
\ee
A 1-cochain is a local integral of the metric, its inverse and their derivatives, linear in $\omega$, say $\Delta[\omega,g]$. A 1-cocycle is a 1-cochain satisfying\footnote{In the commutative formulation of cohomology condition \eqref{cocycle} is replaced by, see \cite{bonoracottareina1983},
\be
\delta_{\omega_1} \Delta[\omega_2,g]- \delta_{\omega_2} \Delta[\omega_1,g]=0\0
\ee
where $\omega_1,\omega_2$ are two different local Weyl transformation parameters. Needless to say condition \eqref{cocycle} is much simpler to handle.}
\be
\delta_\omega \Delta[\omega,g]=0\label{cocycle}
\ee
A cocycle is non-trivial (a true anomaly) if it cannot be written as $\delta_\omega C[g]$ for any 
local 0-cochain $C[g]$, i.e. for any local integral of the metric, its inverse and their derivatives. A trivial cocycle is also called a coboundary.

In the following we understand that diffeomorphisms are a symmetry of the theory under scrutiny. Thus we shall consider only diffeomorphism invariant functionals. For this to work for functional of the type $\Delta[\omega,g]$ it is understood that, under a diffeomorphism represented by the local parameter $\xi^\mu$, $\omega$ transforms as
\be
\delta_\xi \omega = \xi^\mu \partial_\mu \omega\0
\ee 
The basis of our analysis will be the following standard Weyl transformation formulas with $\sfd=6$
\be
\delta_\omega R_{\mu\nu\lambda\rho}& =& 2\omega\, R_{\mu\nu\lambda\rho}-D_\mu D_\lambda \omega\,g_{\nu\rho} + D_\nu D_\lambda \omega\, g_{\mu\rho}+D_\mu D_\rho \omega\, g_{\nu\lambda}-D_\nu D_\rho \omega\, g_{\mu\lambda}\0\\
\delta_\omega R_{\mu\nu} &=&(2-\sfd) D_\mu D_\nu \omega -\square \omega\, g_{\mu\nu}\0\\
\delta_\omega R&=&2(1-\sfd) \square\omega -2\omega\, R\0\\
\delta_\omega \sqrt{g} &=& \sfd\, \omega\, \sqrt{g}\0\\
\delta_\omega \Phi&=& \square \delta_\omega \Phi -2\omega \, \square \Phi +(\sfd -2) D^\lambda \omega\, D_\lambda \Phi\label{Weyltransf}
\ee
for any scalar $\Phi$.

Now let us start the analysis. Auxiliary material is collected in Appendix.

\section{Weyl 1-cocycles in $\sfd$=6}
 
In $\sfd=6$ the most general diffeomorphism invariant 1-cochain of canonical dimension 6 is the superposition of 17 different elements
\be
\Delta[\omega,g] = \sum_{i=1}^{17} a_i \Delta^{(i)}[\omega,g],\quad\quad
   \Delta^{(i)}[\omega,g]= \int d^6x \sqrt{g} \, \omega\, K_i[g]\label{6Deltaomegag}
\ee
where $K_i[g]$ are the following covariant expressions numbered from 1 to 17:
\be
&&R^3, \quad\quad RR_{\mu\nu}R^{\mu\nu},\quad\quad RR_{\mu\nu\lambda\rho}R^{\mu\nu\lambda\rho},\quad\quad R_{\mu\nu}R^{\nu\lambda}R_{\lambda }{}^\mu, \quad\quad R_{\mu\nu}R^{\mu\lambda\rho\nu}R_{\lambda \rho}\0\\
&&R_{\mu\nu} R^{\mu\lambda \rho \sigma} R^\nu{}_{\lambda\rho\sigma},\quad\quad R_{\mu\nu\lambda\rho} R^{\lambda \rho \sigma\tau} R^{\mu\nu}{}_{\sigma\tau},\quad\quad R_{\mu\nu\lambda\rho} R^{\tau\nu\lambda\sigma} R^\mu{}_{\tau\sigma}{}^\rho,\quad\quad R\square R\0\\
&& R_{\mu\nu}\square  R^{\mu\nu},\quad\quad  R^{\mu\lambda \rho \sigma}\square R_{\mu\lambda \rho \sigma},\quad\quad R_{\mu\nu}D^\mu D^\nu R,\quad\quad D^\rho R^{\mu\nu} D_\rho R_{\mu\nu}\0\\
&&D^\rho R^{\mu\nu} D_\mu R_{\nu\rho},\quad\quad D^\mu  R^{\nu\lambda \rho \sigma} D_\mu  R_{\nu\lambda \rho \sigma} ,\quad\quad\square R^2 ,\quad\quad \square^2 R\label{6dcochains}
\ee
and $a_i$ are numerical coefficients.

When applying $\delta_\omega$ to any $\Delta^{(i)}[\omega,g]$ we get a sum of terms quadratic in $\omega$. We will express all of them in terms of a basis of nine independent 2-cochains $\Omega^{(k)}[\omega,g] = \int d^6x\, \sqrt{g} H^{(k)}[\omega,g]$, with $k=1,\ldots,9$. The $ H^{(k)}[\omega,g]$ are ordered from 1 to 9 in the following list\footnote{Due to a misprint the last term was missing in \cite{bonorabregolapasti1986}.} 
\be
&&R^2 \omega\square \omega,\quad\quad R_{\mu\nu}R^{\mu\nu} \omega\square \omega, \quad\quad RR_{\mu\nu}\,\omega D^\mu D^\nu \omega,\quad\quad R^{\mu\lambda}R_\lambda{}^\nu \omega D_\mu D_\nu \omega\label{6d2cochains}\\
&& R^{\tau\nu\lambda\sigma}R_{\tau\nu\lambda\sigma}\omega\square \omega,\quad \square R\,\omega\square \omega,\quad R\, \omega\square ^2\omega,\quad R^{\mu\nu} \omega D_\mu D_\nu\square \omega,\quad \omega D_\mu D_\nu \omega R^{\mu\lambda\rho\nu}R_{\lambda\rho} \0
\ee
The consistency conditions can be written as follows
\be
0= \delta_\omega \Delta^{(i)}[\omega,g] = \sum_{k=1}^9 f_k(a_i) \Omega^{(k)}[\omega,g]\label{6dconsistencyconditions}
\ee
where $f_k(a_i)$ are numerical linear combinations of the coefficients $a_i$. The action of $\delta_\omega $ on the $\Delta^{(i)}[\omega,g]$'s is
\be
\delta_\omega\Delta^{(1)}&=& 30\, \Omega^{(1)}\label{deltaDelta}\\
\delta_\omega\Delta^{(2)}&=& 2 \,\Omega^{(1)}+10\, \Omega^{(2)}+8\Omega^{(3)}\0\\
\delta_\omega\Delta^{(3)}&=& 8\, \Omega^{(3)}+10\, \Omega^{(5)}\0\\
\delta_\omega\Delta^{(4)}&=& 3\, \Omega^{(2)}+12\, \Omega^{(4)}\0\\
\delta_\omega\Delta^{(5)}&=&- 2\, \Omega^{(2)}-2\, \Omega^{(3)}+2\,\Omega^{(4)}+8\, \Omega^{(9)}\0\\
\delta_\omega\Delta^{(6)}&=& \Omega^{(1)}-4\, \Omega^{(2)}-4\,\Omega^{(3)}+12\, \Omega^{(4)}+2\, \Omega^{(5)}-12 \,\Omega^{(9)}\0\\
\delta_\omega\Delta^{(7)}&=&3\, \Omega^{(1)}-12 \,\Omega^{(2)}-12\,\Omega^{(3)}+24\, \Omega^{(4)}+3\, \Omega^{(5)}-24\, \Omega^{(9)}\0\\
\delta_\omega\Delta^{(8)}&=&\frac 34\, \Omega^{(1)}-3 \,\Omega^{(2)}-3\,\Omega^{(3)}+6\, \Omega^{(4)}+\frac 34\, \Omega^{(5)}\0\\
\delta_\omega\Delta^{(9)}&=& 2\, \Omega^{(1)}+10\, \Omega^{(6)}+10\, \Omega^{(7)}\0\\
\delta_\omega\Delta^{(10)}&=&- \Omega^{(1)}+6\, \Omega^{(2)}+4\,\Omega^{(3)}+4\Omega^{(4)}+3\, \Omega^{(6)}+\, \Omega^{(7)}+4 \,\Omega^{(8)}+16\,\Omega^{(9)}\0
\ee
\be
\delta_\omega\Delta^{(11)}&=&4\Omega^{(1)}+12\, \Omega^{(2)}+16\,\Omega^{(3)}-16\Omega^{(4)}+4\, \Omega^{(5)}+2\, \Omega^{(6)}+4 \,\Omega^{(8)}+32\,\Omega^{(9)}\0\\
\delta_\omega\Delta^{(12)}&=& \,\Omega^{(1)}+5\, \Omega^{(6)}+10\Omega^{(8)}\0\\
\delta_\omega\Delta^{(13)}&=&\Omega^{(1)}-6\, \Omega^{(2)}-4\,\Omega^{(3)}-4\Omega^{(4)}-3\, \Omega^{(6)}+\, \Omega^{(7)}-8 \,\Omega^{(8)}-16\,\Omega^{(9)}\0\\
\delta_\omega\Delta^{(14)}&=&-\frac 12 \,\Omega^{(1)}-\, \Omega^{(2)}+2\,\Omega^{(3)}-14\Omega^{(4)}-\frac 52\, \Omega^{(5)}+\frac 32\, \Omega^{(7)}-8 \,\Omega^{(8)}-8\,\Omega^{(9)}\0\\
\delta_\omega\Delta^{(15)}&=&4\Omega^{(1)}-12\, \Omega^{(2)}-16\,\Omega^{(3)}+16\Omega^{(4)}-4\, \Omega^{(5)}-2\, \Omega^{(6)}+2 \,\Omega^{(7)}-8 \,\Omega^{(8)}-32\,\Omega^{(9)}\0\\
\delta_\omega\Delta^{(16)}&=&0\0\\
\delta_\omega\Delta^{(17)}&=&0\0
\ee

 Now, eqs.\eqref{6dconsistencyconditions} are a linear system of 9 equations in 17 unknowns $a_i$.  The coefficients in the RHS's of \eqref{deltaDelta} define a matrix of rank 7. Thus the number of cocycles is 10. Now we have to determine which are the coboundaries. To this end we must determine the most general Weyl invariant cochains (or 0-cocycles).

\section{0-cocycles in 6$\sfd$}

We start with the most general diffeomorphism invariant with six canonical dimensions. This is
\be
C[g]=\sum_{i=0}^{11} b_i C^{(i)}[g],\quad\quad   C^{(i)}[g]=\int d^6x\, \sqrt{g} K^{(i)}[g]  \label{6dinvariant}
\ee
These $K^{(i)}[g]$ are the first eleven expressions in the list \eqref{6dcochains} (the remaining 6 are easily seen to reduce upon partial integration to the first 10 or a combination thereof). The invariance condition is written
\be
0=\delta_\omega C[g]= \sum_{k=1}^{17} f'_k(b_i) \Delta^{(k)}[\omega,g] \label{6dinvariance}
\ee
where $f'_k(b_i)$ are numerical linear combinations of the unknowns $b_i$. The action of $\delta_\omega$ on the $ C^{(i)}[g]$ is given by
\be
\delta_\omega  C^{(1)}&=&-30 \,\Delta^{(16)}\label{deltaC}\\
\delta_\omega  C^{(2)}&=&4\, \Delta^{(9)}-20\, \Delta^{(10)}-8\, \Delta^{(12)}-20\, \Delta^{(13)} -6 \, \Delta^{(16)}\0\\
\delta_\omega  C^{(3)}&=&4\, \Delta^{(9)}-20\, \Delta^{(11)}-8\, \Delta^{(12)}-20\, \Delta^{(13)} -4 \, \Delta^{(16)}\0\\
\delta_\omega  C^{(4)}&=&-12 \, \Delta^{(4)}-12 \, \Delta^{(5)} +3\Delta^{(9)}-6 \, \Delta^{(10)}-12\, \Delta^{(12)}-6\, \Delta^{(13)}-12\, \Delta^{(14)} -\frac 32 \, \Delta^{(16)}\0\\
\delta_\omega  C^{(5)}&=&-10 \, \Delta^{(4)}-10 \, \Delta^{(5)}-4  \, \Delta^{(6)}+2\, \Delta^{(7)}-8 \, \Delta^{(8)}-\frac 12\,\Delta^{(9)}+12 \, \Delta^{(10)}+2  \, \Delta^{(11)}\0\\
&&-4\, \Delta^{(12)}+20\, \Delta^{(13)}-18\, \Delta^{(14)} +\frac 34 \, \Delta^{(16)}\0\\
\delta_\omega  C^{(6)}&=&6 \, \Delta^{(6)}-3 \, \Delta^{(7)} +12\Delta^{(8)}+ \, \Delta^{(9)}-4\, \Delta^{(10)}-7  \, \Delta^{(11)}-2 \, \Delta^{(12)}-16\, \Delta^{(13)}\0\\
&&-12\, \Delta^{(14)} -4 \, \Delta^{(15)}\0\\
\delta_\omega  C^{(7)}&=&12 \, \Delta^{(6)}-6\, \Delta^{(7)} +24\Delta^{(8)}-12 \, \Delta^{(11)}-24\, \Delta^{(13)}+24\, \Delta^{(14)}-6\, \Delta^{(15)}\0
\ee
\be
\delta_\omega  C^{(8)}&=&-6 \, \Delta^{(4)}-6 \, \Delta^{(5)}+6 \, \Delta^{(10)}-\frac 32\,\Delta^{(11)}-3 \, \Delta^{(12)}+6  \, \Delta^{(16)}-6\, \Delta^{(14)}-\frac 32\, \Delta^{(15)}\0\\
\delta_\omega  C^{(9)}&=&-2\, \Delta^{(16)}-20\, \Delta^{(17)}\0\\
\delta_\omega  C^{(10)}&=&-20 \, \Delta^{(4)}-20 \, \Delta^{(5)}-8  \, \Delta^{(6)}+4\, \Delta^{(7)}-16 \, \Delta^{(8)}+3\,\Delta^{(9)}+4 \, \Delta^{(10)}+4  \, \Delta^{(11)}\0\\
&&-16\, \Delta^{(12)}+20\, \Delta^{(13)}-36\, \Delta^{(14)} -\frac 32 \, \Delta^{(16)}-6\, \Delta^{(17)}\0\\
\delta_\omega  C^{(11)}&=&-16 \, \Delta^{(4)}-16\, \Delta^{(5)}-16\, \Delta^{(6)} +8 \, \Delta^{(7)}-32  \, \Delta^{(8)}+4\, \Delta^{(9)}+8\,\Delta^{(10)} \0\\
&&-16 \, \Delta^{(12)}+40\,\Delta^{(13)}-48 \, \Delta^{(14)}-8  \, \Delta^{(15)}-2\, \Delta^{(16)}-4\, \Delta^{(17)}\0
\ee
The system of equations \eqref{6dinvariance} in the unknown $b_i$, characterized by the matrix of coefficients in the RHS's of \eqref{deltaC}, has rank 6, which means that we have five invariants. Therefore we are left with 6 coboundaries. In total we have therefore $10-6=4$ non-trivial cocycles.

\section{Results: non-trivial 1-cocycles and invariants}

The four non-trivial 1-cocycles can be chosen to be
\be
\EA_i[\omega,g] = \int d^6x \, \sqrt{g} \,\omega \, M_i[g], \quad\quad i=1,\ldots,4\label{6dnontrivialcocycles}
\ee
where\footnote{The $M_i[g]$ are the same as in \cite{bonorabregolapasti1986}, except for the coefficient of the first term in $M_5[g]$ and for an unconventional notation where the Ricci tensor and scalar were replaced by their opposite.}
\be
M_1[g]\!\!\!&=&\!\!\! W_{\mu\nu\lambda\rho} W^{\tau\nu\lambda\sigma} W^\mu{}_{\tau\sigma }{}^\rho= \frac{19}{800}K_1-\frac {57}
{160} K_2 +\frac 3{40} K_3 +\frac 7{16}K_4 -\frac 98 K_5 -\frac 34 K_6 +K_8\0\\
M_2[g] \!\!\!&=&\!\!\!W_{\mu\nu\lambda\rho} W^{\lambda\rho\sigma\tau} W_{\sigma\tau }{}^{\mu\nu}= \frac{9}{200}K_1-\frac {27}{40} K_2 +\frac 3{10} K_3 +\frac 5{4}K_4 -\frac 32 K_5 -3 K_6 +K_7 \0\\
M_3[g] \!\!\!&=&\!\!\!K_1-12 K_2+3K_3+16K_4 -24K_5 -24 K_6 +4K_7 +8 K_8\0\\
 M_4[g] \!\!\! &=& \!\!\! -\frac 13 K_1 -8K_2-2K_3 +10K_4-10 K_5 +\frac 12 K_9 -5 K_{10}+5 K_{11}
\label{6dMi}
\ee 
The tensor $ W_{\mu\nu\lambda\rho} $ is the Weyl tensor in six dimensions
 and $M_3[g]$ corresponds to the Euler density. The expressions
\be 
{\cal I}_i[g] = \int d^6x \sqrt{g} M_i[g], \quad\quad i=1,2,3\label{calIi}
\ee
are Weyl invariants. The other two Weyl invariants contain derivatives of the Riemann tensor or the Ricci tensor and scalar.

Let us be more explicit. The most general invariant solutions can be represented by the following values of the coefficients $b_i$:
\be
&&b_6= -\frac{375}2 b_1 -\frac {119}4 b_2 -\frac {27}2 b_3 -\frac {33}8 b_4 +\frac {29}8 b_5\0\\
&&b_7=\frac {575}4 b_1 +\frac {175}8 b_2 +\frac {35}4 b_3 +\frac {41}{16} b_4 - \frac {37}{16} b_5\0\\
&& b_8= - 175 b_1 - \frac {47}2 b_2 -11 b_3 - \frac 94 b_4 +\frac 54 b_5\0\\
&&b_9= -15 b_1-2 b_2 -b_3 + \frac 14 b_5\0\\
&&b_{10} = \frac{75}2 b_1 + \frac {19}4 b_2 +\frac 72 b_3 -\frac 38 b_4 - \frac 58 b_5\0\\
&&b_{11}= \frac {75}4 b_1 +\frac {23}8 b_2 -\frac 14 b_3 +\frac 9{16}b_4 - \frac 5{16}b_5\label{x6-x11}
\ee
while $b_1,\ldots, b_5$ are arbitrary. The Weyl invariants \eqref{calIi} are obtained by imposing that $b_9=b_{10}=b_{11}=0$. Therefore the other two independent Weyl invariants must contain terms with explicit derivatives.

The most general solutions of \eqref{6dconsistencyconditions} in terms of the coefficients $a_i$ are
\be
a_8&=& 8 a_2+8 a_3+4 a_4-4 a_5+4 a_6+4 a_7,\0\\
a_9&=& 3 a_1+\frac{87 a_5}{100}-\frac{37 a_2}{25}-\frac{107 a_3}{25}-\frac{51 a_4}{50}-\frac{93 a_7}{50}-\frac{153 a_6}{100},\0\\
a_{11}&=& -15 a_1+\frac{39 a_2}{10}+\frac{77 a_3}{5}+\frac{57 a_4}{20}+\frac{103 a_6}{20}+\frac{63 a_7}{10}-a_{10}-\frac{57 a_5}{20},\0\\
a_{12}&=& -6 a_1+\frac{79 a_2}{25}+\frac{194 a_3}{25}+\frac{117 a_4}{50}+\frac{69 a_6}{25}+\frac{78 a_7}{25}-\frac{77 a_5}{50},\0\\
a_{13}&=& -\frac{13 a_2}{2}-14 a_3+\frac{15 a_5}{4}+a_{10}-\frac{15 a_4}{4}-\frac{33 a_7}{4}-\frac{49 a_6}{8},\0\\
a_{14}&=& 7 a_2+12 a_3+\frac{9 a_4}{2}+\frac{23 a_6}{4}+\frac{15 a_7}{2}-\frac{7 a_5}{2},\0\\
a_{15}&=& -15 a_1+\frac{27 a_2}{5}+\frac{97 a_3}{5}+\frac{18 a_4}{5}+\frac{32 a_6}{5}+\frac{39 a_7}{5}-a_{10}-\frac{18 a_5}{5}\label{a8-a15}
\ee
while $a_1,\ldots, a_8$ and $a_{10}$ are arbitrary. The first three solutions \eqref{6dnontrivialcocycles} are obtained by imposing $a_9=a_{10}=\ldots=a_{17}=0$. The values \eqref{a8-a15} define 8 solutions.
It is easily seen that $ \Delta^{(16)}[\omega,g] $ and $ \Delta^{(17)}[\omega,g]$ are also solutions of  \eqref{6dconsistencyconditions}, but they are coboundaries. So, in fact, we have altogether 10 solutions of eq.\eqref{6dconsistencyconditions} and 6 coboundaries. The four nontrivial cocycles are recorded in \eqref{6dnontrivialcocycles}. Let us remark that $M_4[g]$ contains explicit derivatives of the Riemann tensor and the Ricci tensor and scalar. It is easy to prove that the four cochains ${\cal A}_i[\omega,g]$ , $i=1,2,3,4$ are indeed cocycles: they cannot be reproduced by acting with $\delta_\omega$ on $C[g]$ for any value of the coefficients $b_i$.

\subsection*{Appendix. Remarkable identities}

Using the properties of the Riemann tensor one can derive the following formulas, which have been used throughout:
\be
D^\mu R_{\mu\nu}&=& \frac 12 D_\nu R\0\\
D^\alpha D^[\mu R_{\alpha\mu\lambda\rho} &=& D_\lambda R_{\mu\rho} -D_\rho R_{\mu\lambda}\0\\
D^\mu R^{\nu\lambda\rho\sigma} D_\nu R_{\mu\lambda\rho\sigma}&=& \frac 12 D^\mu R^{\nu\lambda\rho\sigma} D_\mu R_{\nu\lambda\rho\sigma}\0\\
D^\mu R^{\nu\lambda\rho\sigma} D_\mu R_{\nu\rho\lambda\sigma}&=& \frac 12 D^\mu R^{\nu\lambda\rho\sigma} D_\mu R_{\nu\lambda\rho\sigma}\0\\
R^{\mu\alpha} R_{\alpha\nu\sigma\tau}R_{\mu}{}^{\sigma\nu\tau}&=&\frac 12 R^{\mu\alpha} R_{\alpha\nu\sigma\tau}R_{\mu}{}^{\nu\sigma\tau}\0\\
R^{\rho\mu\tau\alpha}R_{\rho\alpha\sigma\nu} R_{\mu\tau}{}^{\sigma\nu}&=&- \frac 12 R^{\rho\alpha\mu\tau} R_{\rho\alpha\sigma\nu} R_{\mu\tau}{}^{\sigma\nu}\0\\
D^\mu D^\nu \square R_{\mu\nu}&=& 4D^\mu R^{\nu\lambda} D_{\lambda} R_{\mu\nu} -3 D^\lambda R^{\mu\nu}D_\lambda R_{\mu\nu} +2 R^{\mu\nu} R_{\nu\lambda}R^{\lambda}_\mu+2 R^{\nu\sigma} R_{\mu\nu\sigma\lambda}R^{\mu\lambda} \0\\
&&\!\!\!\!\!\!\!\!\!\!\!+2 D^\mu D^\nu R\, R_{\mu\nu}+2 R^{\mu\sigma\nu \lambda} D_\mu D_\lambda R_{\sigma\nu}  -R^{\mu\nu} \square R_{\mu\nu} +\frac 12 \square^2 R+\frac 14 \square R^2 -\frac 12 R\square R\0\\
\square D_\lambda \Phi &=&D_\lambda \square \Phi +R_\lambda{}^\rho D_\rho \Phi\0
\ee
for any scalar $\Phi$.

From these and \eqref{Weyltransf} we can derive the following remarkable identities:
\be
\int d^\sfd x \, \sqrt{g} \,\omega \, R^{\mu\lambda\rho\nu} D_\mu D_\nu R_{\lambda\rho}=\frac 12 \Delta^{(6)} -\frac 14 \Delta^{(7)} +\Delta^{(8)} -\frac 14 \Delta^{(11)}\0
\ee
and
\be
\!\!\!\!\!\!\int d^6x \sqrt{g}\, \omega D_\mu D_\nu\,\omega \, R^{\mu\lambda\rho\sigma} R^\nu{}_{\lambda\rho\sigma}&=& \frac 14 \Omega^{(1)} -\Omega^{(2)} -\Omega^{(3)}+2 \Omega^{(4)}  +\frac 14 \Omega^{(5)}-2 \Omega^{(9)}\0\\
\int d^6x \sqrt{g}\, \omega D_\mu D_\nu\omega\, \square R_{\mu\nu}& =& -\frac 14  \Omega^{(1)} +\frac 12\Omega^{(2)} +\Omega^{(3)} +\frac 12 \Omega^{(6)}+2 \Omega^{(9)}\0\\
\int d^6x \sqrt{g}\, \omega D_\mu D_\nu D_{\lambda} \omega \,D^\mu R^{\nu\lambda}& =& \frac 14  \Omega^{(1)} -\frac 12\Omega^{(2)} -\Omega^{(3)} -\frac 14 \Omega^{(6)}+\frac 14 \Omega^{(7)}-\ \Omega^{(8)}-2 \Omega^{(9)}\0\\
\int d^6x \sqrt{g}\, \omega D_\mu\square \omega \,D^\mu R&=& -\frac 12  \Omega^{(6)}-\frac 12 \Omega^{(7)}\0\\
\int d^6x \sqrt{g}\, \omega D_\mu D_\nu\omega\, D^\mu D^\nu R& =& -\frac 14  \Omega^{(1)} +\Omega^{(3)} + \Omega^{(6)}\0\\
\int d^6x \sqrt{g}\, \omega\square D_\mu D_\nu\omega\, R^{\mu\nu}&=&  -\frac 14  \Omega^{(1)} +\frac 12 \Omega^{(2)} + \Omega^{(3)}+\Omega^{(8)}+2 \Omega^{(9)}\0
\ee

\end{document}